\documentclass[a4paper,10pt]{extarticle}  
\usepackage{graphicx}%
\usepackage{import}
\usepackage{local}
\usepackage[left=2.5cm,top=2cm,bottom=2cm,right=2.5cm]{geometry}
\title{\sffamily\huge\bfseries Soft X-ray imaging with coherence tomography in the water window spectral range using high-harmonic generation}

\author{%
\textbf{Julius Reinhard \orcidlink{0000-0002-5761-4340},\textcolor{Accent}{\textsuperscript{1,2*,\textdagger}} %
Felix Wiesner \orcidlink{0000-0001-5352-9205},\textcolor{Accent}{\textsuperscript{1,2,\textdagger}} %
Themistoklis Sidiropoulos \orcidlink{0000-0003-3305-3750},\textcolor{Accent}{\textsuperscript{3}} %
Martin Hennecke \orcidlink{0000-0003-0826-6471},\textcolor{Accent}{\textsuperscript{3}} %
Sophia Kaleta \orcidlink{0009-0009-7211-2638},\textcolor{Accent}{\textsuperscript{1,2}} %
Julian Späthe \orcidlink{0009-0009-6593-4916},\textcolor{Accent}{\textsuperscript{1}} %
Johann Jakob Abel,\textcolor{Accent}{\textsuperscript{1}} %
Martin Wünsche \orcidlink{0000-0002-1090-8018},\textcolor{Accent}{\textsuperscript{1,2,4}} %
Gabriele Schmidl,\textcolor{Accent}{\textsuperscript{5}} %
Jonathan Plentz \orcidlink{0000-0003-1176-0710},\textcolor{Accent}{\textsuperscript{5}} %
Uwe Hübner,\textcolor{Accent}{\textsuperscript{5}} %
Katharina Freiberg\orcidlink{0000-0001-8821-4529},\textcolor{Accent}{\textsuperscript{6}} %
Jonathan Apell,\textcolor{Accent}{\textsuperscript{6,7}} %
Stephanie Lippmann \orcidlink{0000-0002-8250-4696},\textcolor{Accent}{\textsuperscript{8}} %
Matthias Schnürer,\textcolor{Accent}{\textsuperscript{3}} %
Stefan Eisebitt \orcidlink{0000-0001-7608-5061},\textcolor{Accent}{\textsuperscript{3,9}} %
Gerhard G. Paulus\orcidlink{0000-0002-8343-8811},\textcolor{Accent}{\textsuperscript{1,2}} %
Silvio Fuchs\orcidlink{0000-0002-3246-2627}\textcolor{Accent}{\textsuperscript{1,2,10*}} }\\
\begin{small}\textcolor{Accent}{\textsuperscript{1}}Institute of Optics and Quantum Electronics, Friedrich Schiller University Jena, Max-Wien-Platz 1, Jena, 07743, Germany \\ 
\textcolor{Accent}{\textsuperscript{2}}Helmholtz Institute Jena, GSI Helmholtzzentrum für Schwerionenforschung, Fraunhoferstr. 8, Jena, 07743, Germany \\ 
\textcolor{Accent}{\textsuperscript{3}}Max-Born-Institut für Nichtlineare Optik und Kurzzeitspektroskopie, Max-Born-Straße 2A, Berlin, 12489, Germany \\ 
\textcolor{Accent}{\textsuperscript{4}}Indigo Optical Systems GmbH, Moritz-von-Rohr-Str. 1a, Jena, 07745, Germany \\ 
\textcolor{Accent}{\textsuperscript{5}}Leibniz Institute of Photonic Technology (Leibniz-IPHT), Albert-Einstein-Str. 9, Jena, 07745, Germany \\ 
\textcolor{Accent}{\textsuperscript{6}}Otto Schott Institute of Materials Research Friedrich Schiller University Jena, Löbdergraben 32, Jena, 07743, Germany \\ 
\textcolor{Accent}{\textsuperscript{7}}Institute of Materials Science and Engineering, Chemnitz University of Technology, Erfenschlager Str. 73, Chemnitz, 09125 Germany \\ 
\textcolor{Accent}{\textsuperscript{8}}Institute of Applied Physics, Friedrich Schiller University Jena, Albert-Einstein-Str. 15, Jena, 07745 Germany \\ 
\textcolor{Accent}{\textsuperscript{9}}Institute for Optics and Atomic Physics, Technische Universität Berlin, Straße des 17. Juni 135, Berlin, 10623 Germany\\ 
\textcolor{Accent}{\textsuperscript{10}}Laserinstitut Hochschule Mittweida, University of Applied Science Mittweida, Technikumplatz 17, Mittweida, 09648 Germany \\ 
\textcolor{Accent}{\textsuperscript{\textdagger}}These authors contributed equally to this work.\\
\textcolor{Accent}{\textsuperscript{*}}Correspondence: \textcolor{Accent}{julius.reinhard@uni-jena.de, silvio.fuchs@hs-mittweida.de} \\ \end{small}
}
\date{}
\begin{document}
\maketitle
\thispagestyle{empty}

\section{Abstract}

\noindent
\textbf{\textcolor{Accent}{High-harmonic generation (HHG) is used as a source for various imaging applications in the extreme ultraviolet spectral range. It offers spatially coherent radiation and unique elemental contrast with the potential for attosecond time resolution. The unfavorable efficiency scaling to higher photon energies prevented the imaging application in the soft X-ray range so far. In this work we demonstrate the feasibility of using harmonics for imaging in the water window spectral region (284\,eV to 532\,eV). We achieve nondestructive depth profile imaging in a heterostructure by utilizing a broadband and noise-resistant technique called soft X-ray Coherence Tomography (SXCT) at a high-flux lab-scale HHG source. SXCT is derived from Optical Coherence Tomography, a Fourier based technique, that can use the full bandwidth of the source to reach an axial resolution of 12\,nm in this demonstration. The employed source covers the entire water window, with a photon flux exceeding 10$^\text{6}$\,photons/eV/s at a photon energy of 500\,eV. We show local cross sections of a sample consisting of Aluminium oxide and Platinum layers of varying thickness on a Zinc oxide substrate. We validate the findings with scanning and transmission electron microscopy after preparation with focused ion beam milling.}}

\section{Introduction}
High-harmonic generation (HHG) has proven to be an exceptional source for nanoscale extreme ultraviolet (EUV) imaging due to its laser-like properties and small laboratory footprint \cite{ferray1988multiple,popmintchev2012bright,klas2021ultra}. In particular, coherent imaging methods such as ptychography benefit from the high spatial coherence and have demonstrated their remarkable potential in recent years achieving wavelength-scale resolution \cite{thibault2008high,loetgering2022advances,eschen2022material, gardner2017subwavelength}. In addition, the strong elemental contrast opens up a wide field of applications from semiconductor studies \cite{Tanksalvala2021} and mask inspection \cite{Porter2023} to biological samples \cite{zurch2014cancer,baksh2020quantitative,liu2023visualizing}. Even buried structures can be examined using EUV light from HHG sources \cite{fuchs2016nanometer,fuchs2017optical,shanblatt2016quantitative}. However, the penetration depth is severely limited by the strong absorption of most materials.

Further decreasing the wavelength to the soft X-ray (SXR) range would not only decisively increase the penetration depth for a great variety of materials, but would also enhance the resolution as demonstrated at synchrotron sources \cite{shapiro2020ultrahigh}. Especially the so-called water window (WW), defined by the absorption edges of carbon ($\approx$\,284\,eV or 4.4\,nm) and oxygen ($\approx$\,532\,eV or 2.3\,nm), is of particular interest due to its high contrast and penetration depth in many materials, e.g. biological materials \cite{schneider1998cryo,coale2024nitrogen,reinhard2023laboratory}. 

HHG-driven imaging in the soft X-ray or more specifically the water window spectral range has has not yet been demonstrated, primarily for two reasons. On the one hand, the available photon flux of HHG sources in the SXR range is severely limited. Extending the HHG spectral range into the water window typically requires ultrafast driver lasers with wavelengths above 1\,µm. Unfortunately, this reduces the conversion efficiency of the process, which scales with $\sim\lambda^{-5...-6}$ \cite{tate2007scaling}. Therefore, SXR-HHG have only been used for various spectroscopic applications that exploit the femto- or even attosecond pulse lengths \cite{pertot2017time,attar2017femtosecond,sidiropoulos2021probing}. On the other hand, the intrinsically broadband harmonic spectrum is not efficiently used by the established EUV imaging methods like Ptychography, which typically require monochromatic light and consequently only use a small fraction of the available photon flux. Thus, intrinsically broadband imaging techniques are better suited for flux-limited HHG sources in the water window spectral range.

In this work, we demonstrate the feasibility of HHG-driven imaging in the water window by combining a cutting-edge WW-HHG source with a broadband and noise resistant technique called soft X-ray coherence tomography (SXCT). The HHG source is driven by a few-cycle 2.1\,µm laser \cite{feng202027} and generates a broad spectrum covering the entire water window with more than 5$\times$10$^5$\,ph/eV/s at the oxygen K-edge (2.33\,nm) \cite{van2021high}. To our knowledge, no other HHG source was able to apply these photon numbers in experiments yet. Moreover, SXCT exploits the full bandwidth of the source to reconstruct depth structures with nanometer-scale axial resolution. This imaging technique is derived from spectral-domain optical coherence tomography (OCT, \cite{huang1991optical}) and functions as a reflective method, enabling the investigation of structures on bulk materials such as microchips. It even enables the study of weakly reflective samples due to its Fourier-based nature, which provides exceptional noise resistance. Furthermore, three-dimensional information can be obtained by laterally scanning the sample.

We investigated a sample composed of an aluminum oxide layer and a platinum layer deposited on a zinc oxide substrate using SXCT. A cross-sectional image was obtained scanning across the edge of the buried platinum layer, achieving a depth resolution of 12\,nm. To validate the axial structure and inhomogeneities detected with SXCT, we conducted scanning and transmission electron microscopy (SEM and TEM) following focused ion beam (FIB) milling.
\section{HHG source and SXR Coherence Tomography}
\begin{figure*}[h]
	\centering
	\includegraphics{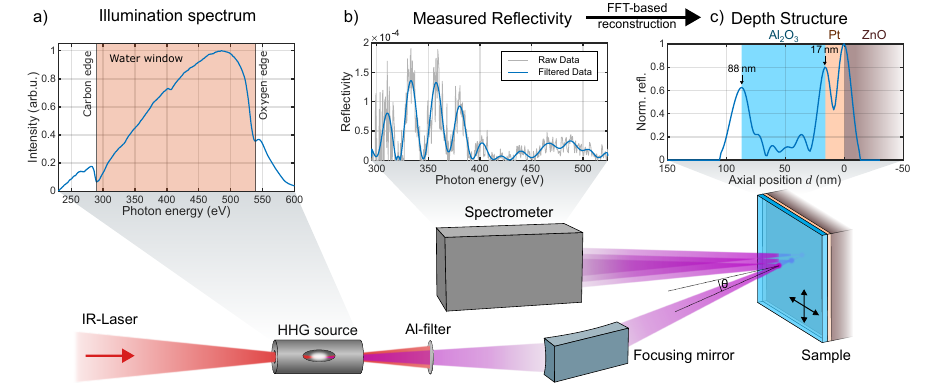}
	\caption{Illumination spectrum, measured sample reflectivity, and reconstructed depth structure arranged on top of a schematic representation of the setup. a) The illumination spectrum spans the entire water window spectral range. It is generated with high-harmonic generation driven by a few-cycle 2.1\,µm laser. The spectrum is shaped by aluminium absorption filters, which contain oxides as well as carbon deposits. These features result in the strongly pronounced absorption edges of oxygen and carbon, which define the water window boundaries. b) The SXR light is focused onto the sample with a toroidal mirror. It is reflected at the various interfaces within the sample, leading to intensity modulations in the reflected spectrum due to interference. The high roughness of the sample leads to a significant reduction in reflectivity, resulting in a very noisy raw signal (grey). The signal relevant for the reconstruction was filtered out using a low-pass Fourier filter (blue). c) The depth structure of the sample was retrieved using the Fourier based SXCT algorithm. Three interfaces are reconstructed, which correspond to two layers with a thickness of 71\,nm and 17\,nm respectively on top of the substrate. By scanning across the sample, additional lateral information can be obtained.}\label{fig_ovw}
\end{figure*}

The demonstration of HHG-driven water window imaging was enabled by a high-flux HHG source covering the entire water window spectral range \cite{feng202027,van2021high}. The source is based on an optical parametric chirped-pulse amplification (OPCPA) system, driven by a 500\,W thin-disk laser to reach a central wavelength of 2.1\,µm and an average power of 28\,W at pulse duration of 27\,fs. The beam is focused with a 750\,mm lens into a helium gas cell, where the HHG process occurs. The resulting spectrum covers a broad spectral range from 200\,eV to 600\,eV, with a photon flux of 5$\times$10$^5$\,ph/eV/s at the oxygen K-edge at $\approx$\,532\,eV \cite{van2021high}. The HHG spectrum which was used for the SXCT measurements is shown in Fig. \ref{fig_ovw}.a.

Soft X-ray coherence tomography (SXCT) extends optical coherence tomography (OCT) to shorter wavelengths and broader spectral ranges, enabling high-resolution imaging in the soft X-ray regime. The axial resolution $\Delta d$ is independent of the focusing geometry. Instead it is defined by the coherence length $l_c\mathbin{=}\lambda^2/\Delta\lambda$, i.e. a shorter wavelength and broader spectrum lead to higher axial resolution. With commercial OCT systems working in the infrared spectral range, axial resolutions of a few micrometers are achieved \cite{leitgeb2004ultrahigh}. In recent years, XUV coherence tomography (XCT), the extreme ultraviolet descendant of OCT, was established in the silicon transmission window between 30\,eV and 100\,eV \cite{fuchs2017optical, wiesner2021material, wiesner2024optical, abel2024non}. This enabled an axial resolution of $\approx$\,15\,nm, which has been demonstrated in a laboratory setup using HHG \cite{nathanael2019laboratory}.

SXCT is implemented as a common-path Fourier-domain OCT variant \cite{vakhtin2003common,fuchs2016nanometer}, in which the sample's spectral response is detected by a spectrometer without the use of a beam splitter. Instead, the separation of the illumination and reflected beam is achieved through an oblique incidence angle. Consequently, the axial resolution $\Delta d$ additionally depends on the incidence angle $\theta$ relative to the surface normal, following $\Delta d\mathbin{=}\lambda^2/(\Delta\lambda\cos\theta)$. In order to enhance the reflectivity and thereby increase the signal strength, particularly at short wavelengths, larger incidence angles can be utilized. However, this results in a slightly reduced axial resolution.  In this experiment, we used an incidence angle of $\theta$\,=\,72$^\circ$, enabling a maximum resolution of $\approx$\,10\,nm for the spectral range from 284\,eV to 532\,eV.

The setup is schematically depicted in Fig. \ref{fig_ovw}. A toroidal mirror focuses the SXR light from the HHG source onto the sample, forming an elliptical spot of 150\,µm$\,\times\,$300\,µm due to $\theta$\,=\,72$^\circ$. The light reflected from internal sample structures interferes with the light reflected from the surface, which serves as a reference in the common path SXCT scheme. This interference induces modulations in the spectrum of the reflected light, which are detected by a grating spectrometer. By measuring the incident spectrum (Fig. \ref{fig_ovw}.a), the sample reflectivity (Fig. \ref{fig_ovw}.b) can be determined.

The sample's depth structure in the illuminated region is encoded in the spectral modulations of this reflectivity and can be reconstructed using a Fourier-based algorithm. The Fourier transform of the intensity reflectivity of the sample represents the autocorrelation of the axial sample structure, which is intrinsically ambiguous. However, the unambiguous axial structure can be reconstructed from the autocorrelation by employing a phase retrieval algorithm. In this work, we use a three-step one-dimensional phase retrieval algorithm originally developed for XCT \cite{fuchs2017optical}.

Due to the Fourier-based nature of the method, SXCT is highly resistant to noise, similar to the principle of lock-in amplifiers. Individual interfaces within the sample contribute to distinct modulation frequencies in the spectrum. Thus, broad-bandwidth noise can be efficiently filtered out, resulting in a high signal-to-noise ratio, as demonstrated in Fig. \ref{fig_ovw}.b. The measured reflectivity of one sample position is shown in gray, while the reflectivity after low-pass Fourier filtering is shown in blue. Using this processed reflectivity as input, the established XCT reconstruction algorithm extracts the depth profile (Fig. \ref{fig_ovw}.c). Each peak corresponds to an interface in the sample, where the reflections occur.

A cross-sectional or three-dimensional image of the sample can be obtained by laterally scanning in one or two dimensions while retrieving the axial structure point by point. As a consequence, the lateral resolution is independent of the axial resolution and defined by the size of the SXR probe on the sample which in our case was 150\,µm$\,\times\,$300\,µm.

\section{Experimental results}
\subsection{SXCT cross section}
We performed SXCT on a layered test sample, consisting of a $\approx$\,70\,nm thick Al$_2$O$_3$ layer on a ZnO substrate. In a specific region of the sample, an additional Pt layer is buried below the Al$_2$O$_3$ with a thickness of $\approx$\,20\,nm. 

The measured reflectivity of a single lateral sample point is shown in Fig. \ref{fig_ovw}.b, with values in the range of 10$^{-4}$. Compared to simulated data the measured reflectivity is more than one order of magnitude lower \cite{henke1993x}. The deviations can be attributed to differences in material density and non-negligible interface roughness. However, even with such a low reflectivity signal, particularly in comparison to the noise level, our reconstruction algorithm successfully retrieves the depth structure and resolves the position of the three interfaces, indicating a resolution of at least 17\,nm (see Fig. \ref{fig_ovw}.c).

\begin{figure*}[ht]
	\centering
	\includegraphics{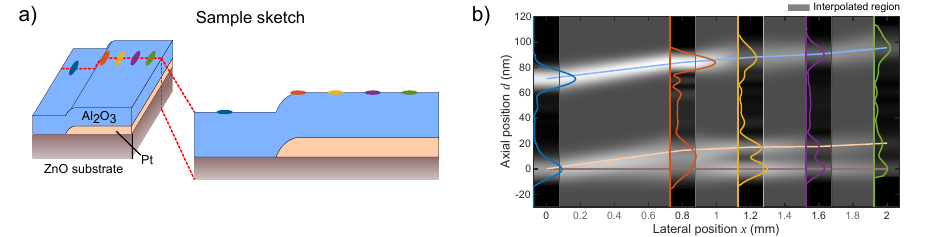}
	\caption{Cross-sectional imaging using SXCT: a) The investigated sample consists of a thin Al$_2$O$_3$ layer on a ZnO substrate, with buried Pt layer present in a specific region of the sample. b) A cross-section was obtained by scanning the sample at five different lateral positions. The size of the SXR-focus was 150\,µm in scanning direction. The grayed-out areas between the scanning positions are interpolated. The Al$_2$O$_3$ layer can be resolved throughout the sample with a varying thickness between 71\,nm and 75\,nm. In addition, the thinner buried Pt structure behind the edge can also be resolved with increasing thickness between 17\,nm and 20\,nm. At the second position (orange) the comparatively large SXR focus averages over a varying Pt thickness, leading to a broadened signal.}\label{fig_results}
\end{figure*}

Subsequently, we generate a cross-section (a so-called B-scan in OCT terminology) by measuring the depth structure (A-scan) at five different lateral positions as indicated with different colors in the schematic illustration of the sample (Fig. \ref{fig_results}.a). Each position required an exposure time of 15 minutes.

The measured cross-section of the sample is presented in Fig. \ref{fig_results}.b. Grayed-shaded areas represent interpolated regions between the scanning positions, which are indicated by the colored graphs and boundaries. Additional horizontal lines track the peak positions, which correspond to the interfaces in the sample. The depth structures were aligned such that the deepest interface (the substrate surface) is located at $d$=\,0\,nm. At lateral position $x$\,=\,0\,mm (blue curve) two peaks are visible: the Al$_2$O$_3$ surface (axial position $d$\,=\,71\,nm) and the Al$_2$O$_3$/ZnO interface ($d$\,=\,0\,nm). As the lateral position $x$ increases, a shift of the surface position becomes apparent. In addition, a broadening of the signal at the Al$_2$O$_3$/Pt interface ($d$\,=\,15\,nm) becomes visible at $x$\,=\,0.8\,mm (orange). This is due to the presence of the buried Pt layer, which apparently does not have a sharp lateral edge, instead exhibiting a gradually increasing thickness. At $x$\,=\,1.2\,mm (yellow) the front ($d$\,=\,17\,nm) and back ($d$\,=\,0\,nm) of the Pt layer are clearly distinguishable. The Pt thickness further increases at $x$\,=\,1.6\,mm (purple) and $x$\,=\,2\,mm (green) reaching 21\,nm. The thickness of the Al$_2$O$_3$-layer increases as well to 75\,nm. Additionally, a decrease in signal strength is observed for larger $x$ values, corresponding to an even lower reflectivity of 10$^{-5}$ for the green signal. This can be attributed to increased roughness, which correlates with the growing Pt layer thickness. The roughness of the Pt layer also influences the roughness of the surface, further reducing the reflectivity.

The peak positions were determined by Gaussian fitting of the depth structure at the scanning positions. This also revealed a FWHM of the Al$_2$O$_3$/Pt interface at the yellow scanning position ($x$\,=\,1.2\,mm, $d$\,=\,17\,nm) of 12\,nm, which defines the achieved resolution. For the purple ($x$\,=\,1.6\,mm) and green scanning position ($x$\,=\,2\,mm) the width increased to 15\,nm, indicating a thickness variation of the Pt layer of $\pm$\,1.5\,nm in the illuminated region.

\subsection{Validation with electron microscopy}
\begin{figure*}[ht]
	\centering
	\includegraphics{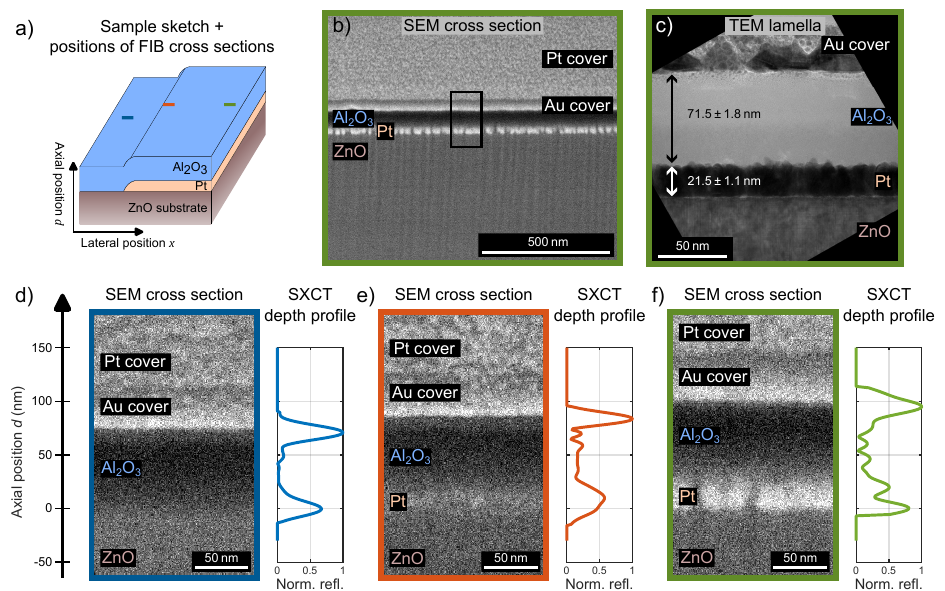}
	\caption{Validation of SXCT results with electron microscopy. a) Three regions on the sample are distinguished: Al$_2$O$_3$ on ZnO substrate (blue position), Al$_2$O$_3$ on Pt on ZnO substrate (green position) and the region in between with the Pt layer slowly forming (orange position). At each position a FIB cross section was made and compared to SXCT depth profiles. b) A SEM image of a FIB cross section at the green position, shows the Al$_2$O$_3$ and Pt layers, as well as the ZnO substrate and additional gold and platinum layers deposited in preparation for FIB milling. The bright Pt layer is easily recognizable and its high granularity and roughness is revealed. c) Section of the TEM lamella, prepared from the cross section shown in (b). The Al$_2$O$_3$ layer is visible on top of the inhomogeneous and rough Pt layer.  Averaging over the full width of 5\,µm, axial distances of 71.5\,nm and 21.5\,nm could be confirmed for Al$_2$O$_3$ and Pt, respectively. d-f) FIB cross sections were made at different sample positions and compared to corresponding SXCT depth profiles. The positions of the interfaces are highly consistent. It is confirmed that the thickness of the Pt layer gradually increases, rather than having a sharp edge.}\label{fig_TEM}
\end{figure*}

We verified the SXCT results by performing scanning and transmission electron microscopy (SEM/TEM) investigations of the sample after FIB preparation. We compared three positions with corresponding SXCT scan positions: Al$_2$O$_3$ on substrate (blue position and blue frame in Fig. \ref{fig_TEM}), Al$_2$O$_3$ with buried Pt (green) and close to the edge (orange). For all three positions, we prepared FIB cross sections and imaged them using SEM. In addition, we extracted a thin lamella with a width of $\approx$\,5\,µm from the Al$_2$O$_3$/Pt position (green) and investigated it with TEM (Fig. \ref{fig_TEM}.c). To protect the sample surface during FIB preparation, the sample was coated with gold and platinum after SXCT measurements. Details can be found in the Methods section.

A large field of view from the FIB cross section at the green position is presented in Fig. \ref{fig_TEM}.b. It can be clearly seen from the granular structure that the platinum grows in columns and does not form a homogeneous layer. As a result, the layer exhibits a high roughness. On top of this, the Al$_2$O$_3$/Au interface appears blurred (see also the enlarged area of the SEM image in Fig. \ref{fig_TEM}.f), indicating the transition of roughness from the buried Pt layer to the surface, which further reduces the reflectivity.

The reasons for columnar growth  of platinum can be attributed to two main factors. On the one hand, these are the deposition conditions, specifically the deposition pressure (Ar pressure) and temperature of the substrate surface (Thornton's model \cite{thornton1977high}). On the other hand, it is the growth of platinum on an oxide surface, which can be described as island growth (Volmer-Weber growth mode \cite{volmer1926keimbildung}). This combination results in a highly inhomogeneous vertical growth, particularly in the case of very thin layers. A TEM investigation at the same position (Fig. \ref{fig_TEM}.c) confirms the non-negligible roughness of the Pt layer. 

The lower part of Fig. \ref{fig_TEM} shows SEM images at three different lateral positions, enabling a direct comparison with the corresponding SXCT depth profiles. The SEM images were aligned, such that the ZnO/Al$_2$O$_3$ or ZnO/Pt interface is at $d$\,=\,0. As expected, no Pt is visible in the blue position (Fig. \ref{fig_TEM}.d) and a sharp interface between the dark Al$_2$O$_3$ and the bright gold cover is observed at $d$\,=\,71\,nm, which is in good agreement with the SXCT results. At the orange position (Fig. \ref{fig_TEM}.e), the emergence of the Pt layer on top of the ZnO is visible in both the SEM image and the SXCT profile. The SXCT depth profile however, is integrated over a varying layer thickness due to the lateral extent of the SXR probe (150\,µm$\,\times$\,300\,µm). For this reason, the interfaces cannot be resolved in this transition region of increasing Pt thickness. The enlarged image of the green position (Fig. \ref{fig_TEM}.f), away from the boundary region, shows the fully formed Pt layer. The positions of the front and back of the Pt layer are in good agreement for SEM and SXCT.

From a series of ten measurements along the width of the TEM lamella (Fig. \ref{fig_TEM}.c), a thickness of 21.5$\,\pm\,$1.1\,nm was determined for the platinum layer and 71.5$\,\pm\,$1.8\,nm for the Al$_2$O$_3$ layer. It is important to note that the error reflects the variation of the layer thickness across the width of the TEM lamella, rather than the method's inaccuracy.


\section{Conclusion and Outlook}
In this work, we showed that imaging with high-harmonic generation (HHG) sources is feasible in the water window (WW) spectral range. This was achieved by combinning an HHG source that spans the whole WW range up to the oxygen edge and the flux efficient and noise resistant method of soft X-ray coherence tomography (SXCT). We imaged the internal sample structure of a layer system consisting of Al$_2$O$_3$ and Pt layers with an axial resolution of 12\,nm, which is close to the theoretical limit of 10\,nm for this spectral range. The results of non-destructive SXCT were in remarkable agreement with those obtained using established, but destructive, scanning and transmission electron microscopy (TEM), both in terms of absolute layer thickness (TEM:~21.5\,nm, SXCT:~21\,nm) and thickness variation (TEM:~1.1\,nm, SXCT:~1.5\,nm) for the buried Pt layer. The thickness variation or high roughness and therefore low reflectivity of the test sample demonstrate the method's applicability to real-world samples. For higher-reflectivity samples, even lower axial resolutions in the single-digit nanometer range can be achieved by using steeper angles of incidence. This was demonstrated for a highly reflective periodic sample using an incoherent laser-plasma source, achieving an axial resolution of $<$\,5\,nm \cite{wachulak2018optical}.

Overall, our results represent an important first step towards exploiting the potential of water window harmonics for non-destructive imaging. Soft X-ray coherence tomography (SXCT) offers a valuable approach for studying materials whose absorption is too strong in the extreme ultraviolet (EUV) range. Simultaneously it provides strong elemental contrast, even for low-Z materials, a common challenge encountered in conventional TEM methodologies. In the EUV range, XCT has already been applied to material identification as well as the reconstruction of roughness and layer thicknesses well below the resolution limit \cite{wiesner2021material}. These approach can be extended into the SXR range, enabling further advancements. Cross-sectional imaging with HHG radiation will also facilitate the study of dynamic effects in buried layers with high spatial and femto- or even attosecond temporal resolution\cite{jarecki2024ultrafast}. Additionally, ongoing advancements in ptychographic algorithms aim to leverage broad spectra more effectively \cite{shearer2025robust}. Together with the development of even more powerful HHG sources, these innovations pave the way for future applications in the water window.


\section{Materials and Methods}
\subsection*{Experimental Setup}
The soft X-ray spectra were acquired using a combined $\theta-2\theta$ reflectometry and spectroscopy setup, enabling transmission and reflection geometries for $\theta$ angles of $45^\circ-90^\circ$. The soft X-rays were provided by a laboratory light source based on high-harmonic generation, operated with the noble gas helium ($\approx$\,2.2\,bar gas pressure). The HHG process was driven by 2.1\,µm, 27\,fs (FWHM) infrared pulses generated via optical parametric chirped-pulse amplification (OPCPA). The OPCPA system features a monolithic concept, deriving the seed and pump pulses for the OPA stages from the same 500\,W thin-disk pump laser (TRUMPF Scientific Lasers Dira 500-10), using a 2.1\,µm front-end (Fastlite) for signal generation \cite{feng202027, van2021high}. The system delivers pulses with 10\,kHz repetition rate at an average power of 28\,W, which are focused into a helium gas cell to generate p-polarized $\leq$\,27\,fs soft X-ray pulses covering a continuous spectrum from 200 to 600\,eV. The resulting photon flux is $\approx$\,5$\times$10$^5$\,ph/eV/s @ O K-edge, $\approx$\,10$^6$\,ph/eV/s @ Ti L-edge and $\approx$\,6$\times$10$^6$\,ph/eV/s @ C K-edge, determined using a grating- and CCD-based spectrometer with absolute sensitivity calibration by the German metrology institute (PTB). The broadband pulses are focused onto the sample upon reflection by a toroidal mirror, leading to a footprint of the focal spot of 150\,$\times$\,300\,µm$^2$ (FWHM) for an angle of 72$^\circ$ to the surface normal. After reflection by the sample, the spectrum is dispersed by a reflection variable line spacing (VLS) grating (Hitachi 001-0450, 2400\,l/mm central line density) and recorded on a CCD camera (Greateyes GE 2048 512 BI).

\subsection*{Data analysis}
\begin{figure*}[ht]
	\centering
	\includegraphics{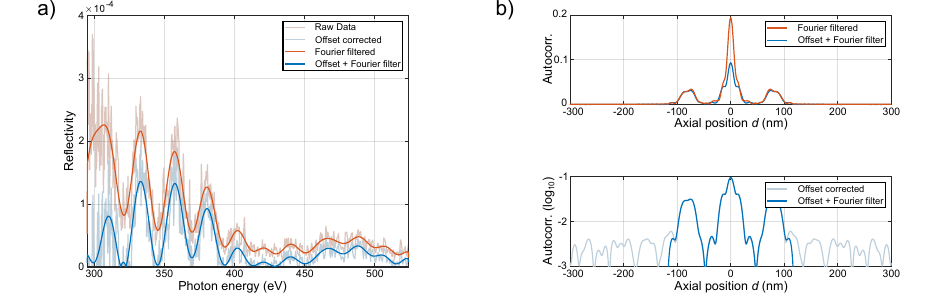}
	\caption{Denoising and offset correction of measured reflectivity data for reconstruction. a) The reflectivity is calculated by dividing the reflected spectrum of the sample by the illumination spectrum. The raw reflectivity gained this way is shown in gray. To denoise the data and eliminate high frequency modulations, a Fourier filter is applied, i.e. all values above or below $\pm$120\,nm in the spatial domain are set to zero (red curve). Additionally, a background correction is applied, by subtracting a constant offset in the wavelength domain (blue). b) Autocorrelation of the sample structure with linear and logarithmic y-axis. The offset correction reduces the signal at zero depth in the autocorrelation as shown in the top figure. The bottom figure displays the applied Fourier filter, eliminating all signal above 120\,nm.}\label{fig:method_offset_denoise}
\end{figure*}

A reference spectrum was recorded prior to the linescan by moving the sample out of the beam path and aligning the spectrometer to the transmitted radiation. Calibration was done by fitting the grating equation to the carbon, nitrogen and oxygen edges visible in the spectra. We then applied offset correction, interpolation into the photon energy domain, and denoising to the recorded spectra, as explained below. The impact of these processing steps on the reflectivity and sample autocorrelation is shown in Fig. \ref{fig:method_offset_denoise}. 

The low reflectivity and incoherently scattered radiation from the inhomogeneous sample cause a significant offset in the measured reflectivity. In the sample autocorrelation this corresponds to an increased signal at zero depth. The phase retrieval relies on the correct correlation between the different features in the sample's autocorrelation. Therefore an offset in the sample reflectivity prevents convergence of the phase retrieval. The effects of the offset correction on the sample reflectivity and autocorrelation are shown in Fig. \ref{fig:method_offset_denoise}.a) and b). In this work, we assume a constant offset on the camera, i.e. in wavelength. We find that applying an offset subtraction of 0.27\,Cts/s yields good results for all points of the linescan except for the point directly at the transition between Pt and Al$_2$O$_3$, where a significantly larger offset is observed and a correction by 0.54\,Cts/s is needed. The reason for this is the laterally inhomogeneous sample within the focal spot. The reflectivity is averagerd over parts with a varying thickness of the Pt layer.

Due to the nonlinear relation between wavelength and photon energy $\lambda \propto 1/E$ the varying energy width of the camera pixels needs to be corrected by dividing by the energy spacing of the grid $\Delta E$. The sample reflectivity is obtained by dividing the reflected spectrum of the sample by the reference spectrum. 

Denoising of the reflectivity is performed by applying a low pass filter in the Fourier domain. This is equivalent to cutting off all signal that lies outside of the sample autocorrelation in depth domain, as can be seen in Fig. \ref{fig:method_offset_denoise}.b). 

The interpolation of the SXCT cross section (Fig. \ref{fig_results}.b) was based on Gaussian curve fitting. For each lateral position Gaussian curves were fitted to the peaks corresponding to the interface positions. The parameters of the curves (position, width and amplitude) were linearly interpolated between the lateral positions and used to plot the cross-section. For the second scan position (orange, $x$\,=\,0.8\,mm) three Gaussians were fitted to describe the broadened peak between $d$\,=\,-10\,nm and $d$\,=\,20\,nm. One Gaussian corresponds to the substrate surface at $d$\,=\,0\,nm, while the other two are used to describe the varying thickness of the Pt layer. The mean value of these two positions is considered as the average position of the Pt/Al$_2$O$_3$ interface at $x$\,=\,0.8\,mm.

\subsection*{Electron microscopy}
Cross sections were prepared using focused ion beam (FIB) milling in an FEI Helios Nanolab 600i dual-beam scanning electron microscope. The sample surface was protected by deposition of protective Au and Pt layers prior to FIB preparation. Cross sections were milled using an ion beam operating at 30\,kV and 2.5\,nA. SEM images were acquired using a secondary electron detector.

Cross section lamellae for transmission electron microscopy (TEM) were prepared using FIB and transferred to a Cu grid before applying the final thinning. Ion beam acceleration voltage and beam current were reduced down to 5\,kV and 15\,pA using stage tilts of 50.5$^\circ$ to 53.5$^\circ$. TEM bright-field imaging was performed using a JEOL NEOARM 200F instrument.

\subsection*{Sample preparation}
The platinum (Pt) and aluminium oxide (Al$_2$O$_3$) layers were deposited on a 10\,$\times$\,10\,mm$^2$ and 1\,mm thick ZnO substrate. The Pt coating was deposited using a sputter coater with Ar as the sputtering gas. The base pressure was 6\,$\times$\,10$^{-2}$\,mbar and the Ar pressure was 2\,$\times$\,10$^{-1}$\,mbar. A rate of approximately 0.14\,nm/s, a voltage of 1.3\,kV and a current of 14\,mA were used.

The deposition of aluminium oxide (Al$_2$O$_3$) has been performed using the atomic layer deposition (ALD) at a temperature of 225$^\circ$C with 500 cycles. Each cycle took 70\,ms. Trimethylaluminium (TMA) was the precursor for the vapor phase deposition.

To create two halves with different layer stacks on the substrate, a mask was used that was removed according to the coating sequence.


\section{Data Availability}
The raw spectra used in this experiment are available from the authors upon reasonable request.

\section{Acknowledgments}
Thanks to Annett Gawlik and Uwe Brueckner from Leibniz IPHT for the deposition of Pt and Al$_2$O$_3$. Funding by ERDF in the project NanoMovie to develop and set up the HHG source at the Max Born Institute is gratefully acknowledged by S.E. This work was supported by the German Research Foundation (grant number PA 730/13-1); European Social Fund (ESF) with Thüringer Aufbaubank (2018FGR008, 2015FGR0094), and Bundesministerium für Bildung und Forschung (VIP “X-CoherenT”).

\section{Author Contributions}
J.R., F.W., S.F. and G.G.P. conceived the experiment. T.S., M.H., M.S. and S.E. conceived the laser system and HHG source and set up the experiment. J.R., F.W., S.K., J.S., T.S. and M.H. conducted the experiment. F.W. and J.R. analysed the data with supervision from S.F. and G.G.P.. The authors F.W., S.F., J.J.A., M.W. and J.R. developed the methodology. G.S., J.P. and U.H. prepared the sample. K.F., J.A. and S.L. performed the FIB milling and electron microscopy measurements. The manuscript was written by J.R., F.W. and S.F. All authors provided critical feedback on the research and the paper.

\section{Author Competing Interests}
The authors declare no competing interests.

\newpage
\renewcommand\refname{References}
\begin{small}
	\bibliographystyle{naturemag.bst} 
	\textnormal{\bibliography{localbibliography.bib}}
\end{small}

\end{document}